# Double carrier transport in electron-doped region in black phosphorus FET

Kohei Hirose[1], Toshihito Osada[1,a)], Kazuhito Uchida[1], Toshihiro Taen[1], Kenji Watanabe[2], Takashi Taniguchi[2], and Yuichi Akahama[3]

[1]*Institute for Solid State Physics, University of Tokyo, Kashiwa, Chiba 277-8581, Japan.*

[2]*National Institute for Materials Science, Tsukuba, Ibaraki 305-0044, Japan.*

[3]*Graduate School of Material Science, University of Hyogo, Kamigori, Hyogo 678-1297, Japan.*

**Abstract**

Double carrier transport has been observed in thin film black phosphorus (BP) field effect transistor (FET) devices in the highly electron-doped region. BP thin films with typical thicknesses of 15 nm were encapsulated by hexagonal boron nitride (h-BN) thin films to avoid degradation by air exposure. Their Hall mobility reached 5,300 cm$^2$/Vs and 5,400 cm$^2$/Vs at 4.2 K in the hole- and electron-doped regions, respectively. The gate voltage dependence of conductivity exhibits an anomalous shoulder structure in the electron-doped region. In addition, at gate voltages above the shoulder, the magnetoresistance changes to positive, and there appears an additional slow Shubnikov-de Haas oscillation. These results strongly suggest the appearance of second carriers, which originate from the second subband with a localized band edge.

________________________________________________________________________

[a)]E-mail: osada@issp.u-tokyo.ac.jp

Black phosphorus (BP) was extensively studied in the 1980s because it is a single-element semiconductor with higher mobility than silicon or germanium.[1-3] It is easy to obtain ultra-thin films of BP by mechanical exfoliation, as BP is a layered crystal with van der Waals interlayer coupling. Since the transport properties of single-layer BP, which is referred to as phosphorene, was reported in 2014,[4] thin film BP has again attracted a great deal of attention as a post-graphene atomic layer material.[5,6] Subsequent transport studies have shown that few-layer BP has high performance as a field effect transistor (FET) material;[7-12] it possesses a considerably larger on/off ratio than graphene and a higher mobility than transition metal dichalcogenides (TMDs).[13]

Because the thin film BP is degraded by photochemical reactions under oxygen and water atmospheres,[14,15] techniques to prevent degradation are important to preserve sample quality. Encapsulation by hexagonal boron nitride (h-BN) thin films is an effective way to prevent air exposure. Using the dry-transfer technique, the so-called stamp method,[16,17] a sandwiched heterostructure, h-BN/BP/h-BN, can be formed on a $SiO_2/n^+$-Si substrate.[18-22] The h-BN layers not only protect BP from degradation, but also improve the sample mobility, because the cleaved surface of an insulating h-BN film is atomically flat, with no dangling bonds.[23] It is also effective to insert a graphite thin film between the encapsulated BP and $SiO_2/n^+$-Si substrate to screen the random potential on the $SiO_2$ surface.[21] The cleaving and transfer processes are often performed in an inert gas atmosphere using a glove box. Electrical contact with the encapsulated BP is formed in various ways, including one-dimensional edge contact,[18,24] area contact on the region where h-BN is partially etched off,[19] and tunneling contact through monolayer h-BN.[20]

The high-quality samples prepared by the above techniques have led to intensive study of the transport properties of thin film BP. The carrier mobility has been improved by one order of magnitude, with the Hall mobility (not the field-effect mobility estimated from the gate characteristics) reaching 6,000 cm$^2$/Vs and 3,400 cm$^2$/Vs at low temperatures for holes and electrons, respectively.[21,22] The high mobility has enabled observance of Shubnikov-de Haas (SdH) oscillations of magnetoresistance (MR) in thin film BP FETs.[18-20,25,26] The SdH oscillations have been primarily investigated in the hole-doped (negatively gated) region with higher mobility, although they appear also in the electron-doped (positively gated) region. In each region, the SdH oscillation shows a single period, spin splitting, and no Berry phase correction. Moreover, the oscillation period is scaled by the normal component of the magnetic field, indicating a two-dimensional (2D) feature of the carriers. This is consistent with the calculated result, that the carriers are confined in approximately two atomic layers in BP FETs.[18,25,26] Subsequently, the integer quantum Hall effect (QHE) has been reported, confirming the 2D feature of the system.[21,22] As described above, so far, single-carrier transport of 2D carriers confined in the inversion layer has been reported in BP-FETs, mostly in the hole-doped (negatively gated) region.

In this paper, we report double carrier transport in BP FETs in the highly electron-doped region. Single crystals of BP were synthesized under high pressure.[27] The encapsulated BP EFT samples were prepared by the standard method mentioned above: the h-BN/BP/h-BN stacking structure is fixed on a SiO$_2$ (300 nm thick) /n$^+$-Si substrate, which is used as a back gate electrode, using mechanical exfoliation and dry transfer techniques in a glove box filled with N$_2$ gas. The typical thickness of a thin film BP flake is 15-20 nm. Here, the electronic contact is made in a much easier way than in previous studies. As shown in Fig. 1(a) and (b), the upper h-BN layer does not cover the entire thin film BP flake, leaving a region of BP uncovered. Voltage contacts of Au/NiCr are formed on BP at the edge of the upper h-BN flake by electron-beam lithography in atmosphere. Although the uncovered part is exposed to air, the measured part of the BP layer is protected from degradation by the upper h-BN layer. The transport measurements of BP EFTs were performed using a 13T superconducting magnet system.

Figure 1(c) shows the gate voltage dependence of conductance in an encapsulated thin film BP FET device at $T = 4.2$ K. The thickness of the thin film BP is 15 nm in this device. The resistance of the covered part of BP is measured by the four-terminal method as a function of back gate voltage, $V_G$. The central part of the data is lacking in the region $-20\text{ V} < V_G < 40\text{ V}$ because the measurement signal was unstable in the device. The uncovered part of BP exhibits a shift in the gate dependence because of carrier doping resulting from the degradation. Because the source and drain contacts (current contacts) were located in the uncovered part in the present sample, the four-terminal resistance measurement became unstable in the gate voltages corresponding to the transport gap ("cutoff" region) of the uncovered BP. Because the missing part includes the whole cutoff region of the covered BP, the present experiments were performed in the "metallic" region, where sufficient carriers are accumulated in the inversion layer around the BP/h-BN interface.

The field-effect mobility is given by $\mu_{FE} = (1/C_G)d\sigma/dV_G$, where $C_G$ and $\sigma$ denote the gate capacitance per unit area and the sheet conductivity, respectively. For the present sample, with a lower h-BN thickness of 15 nm, $C_G$ is estimated to be $1.10\times10^{-8}$ F/cm$^2$ using the dielectric constants of SiO$_2$ (3.9) and h-BN (4.2). Conversely, the Hall mobility is given by $\mu_H = \sigma/n_H e$, where $n_H$ denotes the sheet carrier density determined by the Hall effect. Although the field-effect mobility is easier to evaluate, the Hall mobility reflects the nature of mobile carriers more accurately.

In the hole-doped region, $-80\text{ V} < V_G < -20\text{ V}$, the slope is almost constant, and it gives the field-effect mobility of holes as $\mu_{FE} = 8{,}600$ cm$^2$/Vs. The Hall mobility of holes is estimated to be $\mu_H = 5{,}330$ cm$^2$/Vs at $V_G \sim -80$ V. Conversely, in the electron-doped region, $20\text{ V} < V_G < 100\text{ V}$, an anomalous shoulder structure appears at $V_G > 60$ V. This is the main subject of the present paper. In the electron-doped region below this structure, the field-effect mobility and Hall mobility of electrons are estimated to be $\mu_{FE} = 5{,}100$ cm$^2$/Vs and $\mu_H = 5{,}400$ cm$^2$/Vs, respectively. These values are comparable with the highest reported mobility. Then, we performed four-terminal MR measurements under magnetic fields normal to the layers.

In the hole-doped region, we observed negative MR because of the weak localization of holes and SdH oscillations of holes with clear spin splitting, as shown in Fig. 2. These features are in good agreement with previously observed results, qualitatively: the cyclotron mass is estimated to be $0.2 \sim 0.3\ m_0$ from the temperature dependence (See supplementary material). The sheet carrier density is estimated from the period $\Delta(1/B)$ of the SdH oscillations, using $n_{SdH} = 2e/\{h\Delta(1/B)\}$. Figure 3 shows the dependence of sheet carrier density on the gate voltage. The estimated $n_{SdH}$ values are plotted with solid circles in the negatively gated region. In the "metallic" region, the accumulated carrier density is considered to be determined by the gate capacitance. The solid line indicates the accumulated sheet carrier density expected from the gate voltage, $n_G = C_G V_G/e + n_0$, where $n_0$ is a small correction of charge neutrality. We find that the estimated value is well reproduced by $n_G$ with $n_0 = 3.4 \times 10^{11}$ cm$^{-2}$. This fact means that a simple 2D hole gas with a single closed Fermi surface is formed in the inversion layer.

In contrast, anomalous MR behaviors have been observed in the electron-doped region corresponding to the anomalous shoulder structure above $V_G = 60$ V. With gate voltages below 60 V, normal behaviors, that is, negative MR and SdH oscillations with a single period and spin splitting, were observed, as in the hole-doped region. The cyclotron mass, $m_c^{(1)}$, is estimated to be $0.4 \sim 0.5\, m_0$ from the temperature dependence of the SdH amplitude (See supplementary material). These features are consistent with previously reported results on the electrons in BP FETs.[25] As seen in Fig. 3, the carrier density $n_{SdH}$ estimated from the SdH period (solid circles) below $V_G = 60$ V coincides with $n_G$ (solid line) with the same parameters as the hole-doped region. This fact indicates that a simple 2D electron gas with a single closed Fermi surface exists below $V_G = 60$ V. Here, we note that $n_{SdH}$ is a more reliable carrier density than $n_H$ estimated from the Hall effect, because $n_H$ includes the material-dependent scattering factor $r_H = \langle \tau^2 \rangle / \langle \tau \rangle^2$.

Above $V_G = 60$ V, the transport features change drastically. In the gate voltage dependence at zero magnetic field, the conductance shows a plateau-like saturation in the region $60\text{ V} < V_G < 80\text{ V}$, and increases again at $V_G > 80$ V. The Hall mobility estimated assuming a single carrier decreases in the region $60\text{ V} < V_G < 80\text{ V}$. The MR becomes positive with a steep increase and becomes saturated, accompanied by SdH oscillations, as shown in Fig. 2. Within the framework of the simple Drude model, generally, positive MR cannot occur in a single carrier system with a constant mobility, but it can be obtained in a two-carrier system.[28] The observed large positive MR suggests multi-carrier transport above $V_G = 60$ V. In bulk BP under pressure, two-carrier transport has been observed originating from electron and hole pockets.[29]

In fact, as seen in Fig. 3, the carrier density $n_{SdH}^{(1)}$ estimated from the SdH period (solid circles) slightly deviates from $n_G$ (solid line) above $V_G = 60$ V. This result means that the electrons giving the SdH oscillations are not all the accumulated electrons. Moreover, we can find new slow oscillations at high gate voltages $V_\text{G} > 80\,\text{V}$, as indicated by arrows in Fig. 2. Because this oscillation is periodic with respect to the inverse magnetic field, it can be identified as the second SdH oscillation, which indicates the existence of another Femi surface. Let $n_\text{SdH}^{(2)}$ be the carrier density estimated from the second SdH oscillations. In Fig. 3, the summation $n_\text{SdH}^{(1)} + n_\text{SdH}^{(2)}$ is plotted with solid triangles. We can see that $n_\text{SdH}^{(1)} + n_\text{SdH}^{(2)}$ almost coincides with $n_\text{G}$ (solid line), within reasonable error. This means that there exist only one additional Fermi surfaces, with no valley degeneracy. Despite of error, at least $n_\text{SdH}^{(1)} + n_\text{SdH}^{(2)} = n_\text{G}$ (additional one Fermi surface) is more plausible than $n_\text{SdH}^{(1)} + 2n_\text{SdH}^{(2)} = n_\text{G}$ (additional two Fermi surfaces). Therefore, the additional Fermi surface must be located at a symmetric point of the 2D Brillouin zone because of the inversion symmetry of the system.

In the positively gated BP FETs, the electrons are confined to the inversion layer with a thickness of a few layers, so that the conduction band is quantized into 2D subbands because of the confinement potential. Their electronic structures are expected to be similar to those of few-layer BP. According to band calculations for few-layer BP, there exist several local minima, i.e., valleys, in the conduction band.[6] Except for the band edge at the Γ point, these valleys are located at non-symmetric points in the 2D Brillouin zone, as shown schematically in Fig. 4(a). However, the observed slow SdH oscillations cannot originate from these degenerated valleys, as mentioned above. The second Fermi surface giving the slow oscillations is considered to be that of the second subband at the Γ-point, as shown in Fig. 4(b).

In Fig. 4(b), once the Fermi level $E_\text{F}$ reaches the second subband edge $E_2$ with increasing gate voltage, electrons start to populate the second subband, the edge of which may be localized. When the Fermi level is located above the mobility edge of the second subband, double carrier transport can be observed. At $V_G = 60$ V, just below the shoulder structure, the Fermi energy must be equal to the separation of two subbands, $E_{12}$. Therefore, it can be estimated as $E_{12} = \{\hbar^2/2m_c^{(1)}\}(2\pi n_{SdH}^{(1)}) \sim 24$ meV using the electron density, $n_{SdH}^{(1)} \sim 4.5 \times 10^{12}$ cm$^{-2}$ and the cyclotron mass of the first subband, $m_c^{(1)} = 0.45\ m_0$. Note that this estimated value is for the case of $V_G = 60$ V, as $E_{12}$ is considered to be a function of $V_G$.

In the present study, we measured three h-BN/BP/h-BN samples. We confirmed the appearance of the anomalous shoulder structure in the gate characteristics in all samples. Although fast SdH oscillations were observed in the hole-doped and electron-doped regions in all samples, the slow SdH oscillation was clearly observed only in one sample with the highest mobility."

The appearance of the second carrier at high gate voltages has never been reported. Although Saito and Iwasa injected holes and electrons up to densities of $10^{14}$ cm$^{-2}$ into the inversion layer of BP thin film using the ionic liquid gating technique, they did not observe any sign of the shoulder structure in the gate voltage dependence at $T = 200$ K.[12] This might be because the temperature was too high and the sample quality was not as good. In most experiments using high-quality h-BN/BP/h-BN structures, the hole-doped side has been studied.

In summary, we have observed double carrier transport in thin film BP FETs in the highly electron-doped region ($> 5 \times 10^{12}$ cm$^{-2}$). BP thin films with typical thicknesses of 15 nm were encapsulated by h-BN thin films to prevent degradation by air exposure.

Their Hall mobility reached 5,300 cm$^2$/Vs and 5,400 cm$^2$/Vs at 4.2 K in the hole- and electron-doped regions, respectively. The gate voltage dependence of conductance exhibits an anomalous shoulder structure in the electron-doped region, above which the MR changes to positive and an additional period of SdH oscillations appears. These features suggest that electrons begin to populate the second subband under high gate voltages, resulting in the double carrier transport. The present result demonstrates the feasibility of switching single and double carrier transport using the gate voltage in thin film BP FETs and provides significant information for designing these thin film BP devices.

See supplementary material for analysis of temperature-dependence of SdH oscillations in both hole-doped and electron-doped regions.

**Acknowledgements**

The authors thank Dr. S. Fukuoka and Mr. H. Nakase for valuable discussions and the transfer technique to form the encapsulated BP, respectively. They are also thankful to Prof. H. Tajima for his proposal of the present joint research. This work was supported by JSPS KAKENHI Grant Numbers JP25107003 and JP16H03999.

**Figure 1** (Hirose et al.)

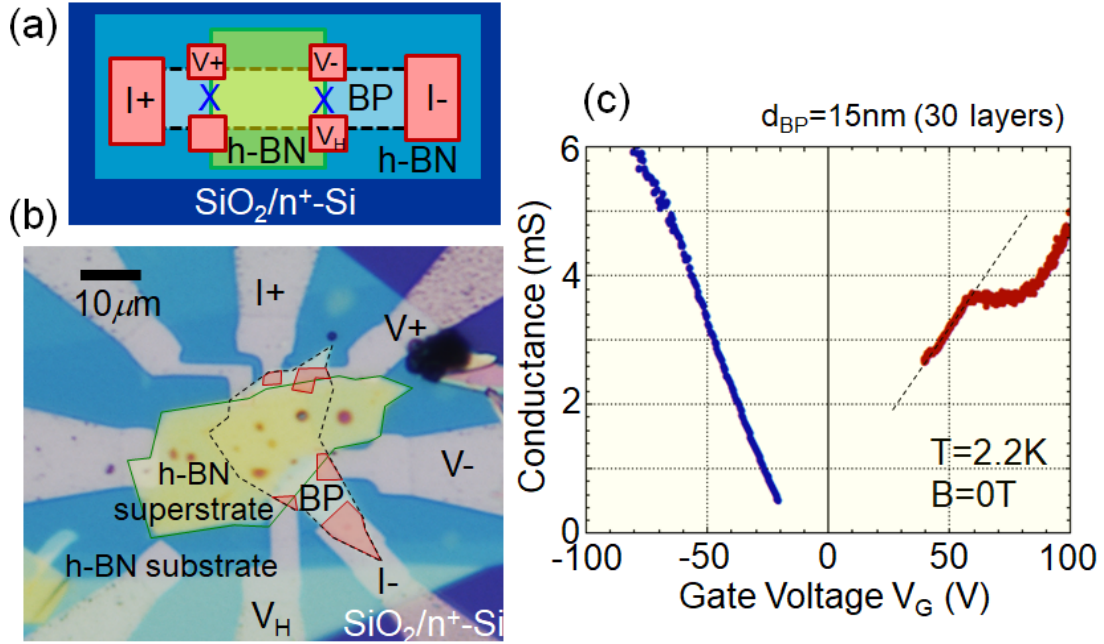

**FIG. 1.** (color online)

(a) Schematic of an h-BN/BP/h-BN device structure. "X" indicates the boundary between high-quality and degraded BP regions. (b) Optical microscope image of an h-BN/BP/h-BN device on a $SiO_2$/n-Si substrate. (c) Four-terminal conductance of an encapsulated thin film BP FET device at $T = 2.2$ K as a function of the back gate voltage. In the region $-20$ V $< V_G <$ 40 V, the measurement was unstable because of high resistance of the degraded BP part. In the highly electron-doped region ($V_G > 60$ V), the conductance shows a shoulder structure.

**Figure 2** (Hirose et al.)

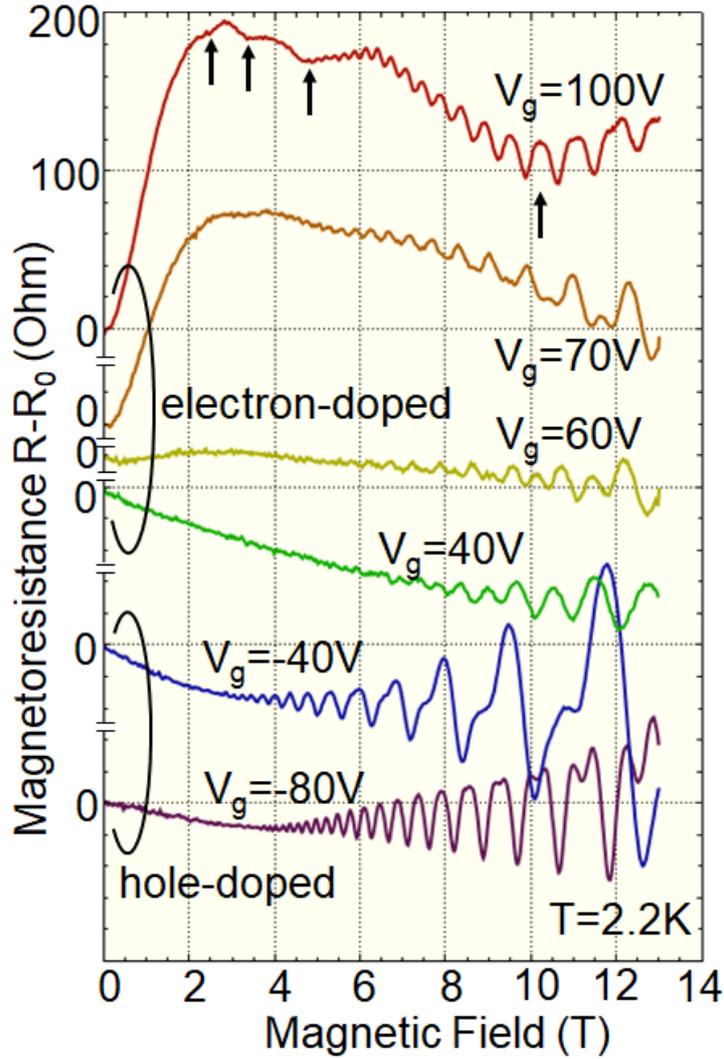

**FIG. 2.** (color online)

Magnetoresistance (MR), which is the change of resistance $R$ from the zero-field resistance $R_0$, in thin film BP FET device measured by four-terminal method at several gate voltages at $T = 2.2$ K. The thickness of the thin film BP is 15 nm (30 layers). In the highly electron-doped region ($V_G > 60$ V), MR becomes positive and new slow oscillations (the second SdH oscillations) appear, as indicated by arrows.

**Figure 3** (Hirose et al.)

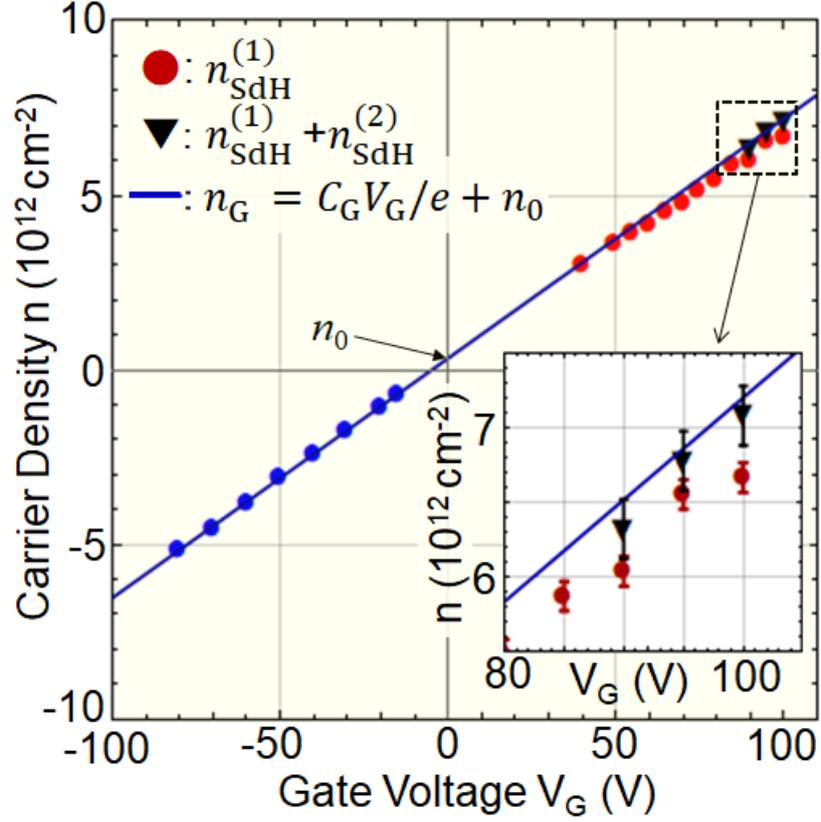

**FIG. 3.** (color online)

Estimated sheet carrier density as a function of gate voltage. Positive/negative carrier density means electron/hole density. The inset shows an enlarged view of the highly electron-doped region. The solid circle indicates the carrier density $n_{\text{SdH}}^{(1)}$ estimated from the rapid SdH oscillation. The solid triangle indicates the summation of $n_{\text{SdH}}^{(1)}$ and $n_{\text{SdH}}^{(2)}$, the latter of which is the electron density estimated from the slow SdH oscillation. The solid line indicates the density estimated from the gate voltage with gate capacitance $C_\text{G} = 1.10 \times 10^{-8}$ F/cm$^2$ and small correction of charge neutrality $n_0 = 3.4 \times 10^{11}$ cm$^{-2}$.

**Figure 4** (Hirose et al.)

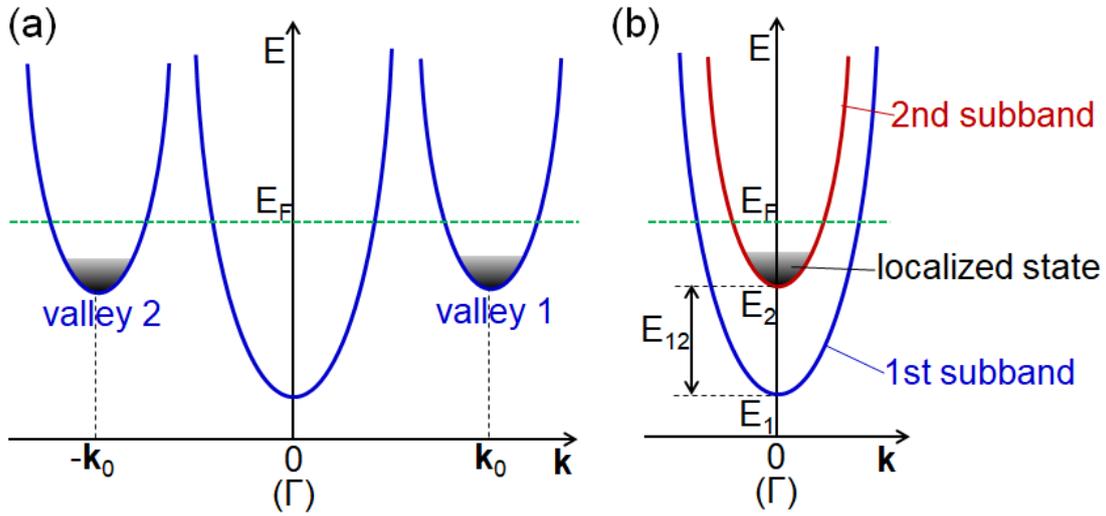

**FIG. 4.** (color online)

Schematic of possible conduction band configurations in 2D **k**-space. (a) Case where electrons populate plural valleys of the conduction band. (b) Case where electrons populate the second subband at **k**=0 (Γ-point). The additional SdH frequency gives the carrier number on the second subband. The electronic states at the second subband edge are localized, with no contribution to transport.